\documentclass[12pt]{article}
    \usepackage{subcaption}

    \usepackage[utf8]{inputenc}
    \usepackage[english]{babel}
    \usepackage[centertags]{amsmath}
    \usepackage{amsmath}
    \usepackage{amssymb}
    \usepackage{bm}
    \usepackage{amsbsy}
    \usepackage{graphicx}
    \usepackage{appendix}
    \usepackage{mathrsfs}
    \usepackage{epsfig}
    \usepackage{color}
    \usepackage{float}
    \usepackage{euscript}
    \usepackage{ulem}
    \usepackage{comment}
    \usepackage{multirow}
    \usepackage{cite}

    \definecolor{dgreen}{rgb}{0,0.6,0}
    \definecolor{darkred}{rgb}{0.8,0,1}
    
    \definecolor{darkblue}{rgb}{0., 0, 1}
    
    \definecolor{darkblue}{rgb}{0., 0, 1}
    
    \definecolor{purple}{rgb}{0.65,0.,0.78}
    
    \definecolor{orange}{rgb}{0.89,0.42,0.05}
    
    \usepackage{jheppubm}

    \newcommand{\be}{\begin{equation}
    }\newcommand{\ee}{
    \end{equation}}
    \newcommand{\bea}{\begin{eqnarray}}
    \newcommand{\eea}{\end{eqnarray}}

    \numberwithin{equation}{section}
    \title{Exact Regular Planar Black Holes solutions in Bumblebee gravity }
    \author{Daniil Stepanenko$^a$}
    \affiliation{$^a$Steklov Mathematical Institute, Russian Academy of
    Sciences, \\ Gubkina str. 8, 119991, Moscow, Russia}
    \emailAdd{dstepanenko@mi-ras.ru}
    \abstract{We discuss possibilities of finding exact regular planar black hole solutions in bumblebee gravity with temporal and radial components of the bumblebee field. We consider quadratic and linear potentials, with vanishing and nonzero cosmological constants, respectively. The field equations determine the nonminimal coupling and the relations between the model parameters. Found solution produce regular planar black hole with planar deformation function. Analytical expressions for this functions has been found. This solutions free of singularities at the $r=0$.  }
    
\begin{document}
    \maketitle
    \section{Introduction}
    \label{sec:introduction}
    
    General relativity and the Standard Model successfully describe gravitational and nongravitational interactions, respectively. Despite their success, integrating GR with the Standard Model remains a major challenge and motivates the study of modified theories of gravity. Although quantum-gravity effects are generally expected near the Planck scale, Lorentz-invariance violation provides a possible low-energy window into the quantum nature of gravity. Originating in string theory, the idea of spontaneous Lorentz-symmetry breaking motivated the Standard-Model Extension (SME), a comprehensive effective-field-theory framework for Lorentz-violating physics \cite{Kostelecky:1988zi,Colladay:1998fq,Kostelecky:2003fs}. A well-motivated realization in the gravitational sector is the bumblebee model, in which a vector field $B_\mu$ acquires a nonzero vacuum expectation value through a symmetry-breaking potential. This vacuum value selects a preferred spacetime direction and naturally incorporates Lorentz violation into the gravitational action \cite{Kostelecky:1989jw,Bertolami:2005bh}. Finding exact black-hole solutions in this model and studying their thermodynamic properties are therefore important for probing the strong-field consequences of Lorentz-symmetry breaking.
    
    The first exact Schwarzschild-like solution was obtained in \cite{Casana:2017jkc}. Other works found Schwarzschild-AdS-like, Kerr-like, BTZ-like, and wormhole solutions \cite{Ovgun:2018xys,Maluf:2020kgf,Ding:2019mal,Xu:2022frb,Ding:2023niy}. Further developments include a Schwarzschild-like bumblebee black hole with a topological defect \cite{Gullu:2020qzu}, extended thermodynamics of bumblebee black holes \cite{Mai:2023ggs}, and spherically symmetric solutions away from the potential minimum \cite{Bailey:2025oun}.
    
    The first charged solution was obtained by Liu et al.\cite{Liu:2024axg}. They obtained first exact charged spherically symmetric of  and slowly rotating charged solutions, both with and without a cosmological constant in the model where electromagnetic field nonminimally coupled to the bumblebee field.
    In a later work \cite{Liu:2025oho}, Liu et al.\  found new  solutions where bumblebee field had temporal and radial components. By studing their thermodynamics it was found that for a lightlike vacuum expectation value, the black-hole entropy agrees with the Wald entropy.
    
    Detailed discussion and derivation of bumblebee black hole thermodynamic in Einstein-Gauss-Bonnet gravity with cosmological constant was perform in \cite{Ding:2021iwv}. Also recent studies of bumblebee black hole  discuss  conserved charges of BTZ-like solutions, phantom nonlinear-electrodynamic rotating solutions and their shadows, primordial-black-hole formation, asymptotically flat black holes with a purely temporal bumblebee field, nonlinear-electrodynamic charged solutions, and the optical and thermodynamic properties of bumblebee AdS black holes \cite{Ditta:2023yig,Ding:2025cno,Li:2025tcd, Sekhmani:2025gvv,Khodadi:2026nja,Li:2026tae,EslamPanah:2026cwt}.
    
    The exact analytic asymptotically flat black-hole solutions and solutions containing naked singularities with their thermodynamics analysis where found in \cite{Yang:2026jey} .
    
    A diverse set of solutions spherically symmetric  solutions with non-minimal couplings, including naked singularities and black holes,  and analyze of their thermodynamical properties was recently  obtained  in \cite{Zhu:2026vae}. 
    The first attempt to obtain black-hole solutions in the Einstein--Bumblebee--dilaton model, including asymptotically Lifshitz solutions, was made in \cite{Lessa:2023yvw,Lessa:2023dbd}. 
    
    As a main result of this paper, we present new regular black hole solutions in bumblebee gravity, motivated by previous studies addressing the resolution of black hole singularities \cite{Bardeen1968,Frolov:1989pf,Frolov:1988vj,Hayward:2005gi,Frolov:2014jva,Frolov:2016pav,Frolov:2017rjz,Carballo-Rubio:2018pmi,Simpson:2018tsi,Simpson:2019mud,Simpson:2021dyo} in various gravitational models, including general relativity coupled to nonlinear electrodynamics \cite{Ayon-Beato:1998hmi,Ayon-Beato:2000mjt,Bronnikov:2000vy}, quasi-topological gravity \cite{Bueno:2024dgm,Frolov:2024hhe,Tsuda:2026xjc}, non-polynomial gravity \cite{Bueno:2025zaj,Colleaux:2026hat,Liu:2026iou}, and models with phantom scalar fields \cite{Bronnikov:2005gm}, as well as recent discussions of regular black holes, their dynamical formation, energy conditions, internal structure, and thermodynamic properties \cite{Lan:2023cvz,Kumar:2024zqf,Carballo-Rubio:2025fnc,Bueno:2024eig,Fernandes:2025eoc,Zhu:2025dur,Shao:2025apr}.
    

    \section{Bumblebee gravity}
    \label{sec:model}
    
    We will consider Bumblebee gravity model with the action
    \begin{equation}
    \label{action}
    S=\int d^{4}x\,\sqrt{-g}\left\{R-2\Lambda+\xi B^{\mu}B^{\nu}R_{\mu\nu}-\frac{1}{4}B_{\mu\nu}B^{\mu\nu}-V\left(B^{\mu}B_{\mu}-b^{2}\right)\right\}.
    \end{equation}
    
    Here, $B_\mu$ is the bumblebee field, $\Lambda$ is the cosmological constant, and $\xi$ is the nonminimal coupling. We define the bumblebee field strength and the constraint function as
    \begin{equation}
    B_{\mu\nu}=\partial_\mu B_\nu-\partial_\nu B_\mu,\qquad X=B^\mu B_\mu-b^2.
    \end{equation}
    
    We will impose  vacuum condition as 
    \begin{equation}
    \label{vacuumcondition}
    V(X)=V(B^\mu B_\mu-b^2)=0, \quad \text{such that} \quad B_\mu B^\mu=b^2.
    \end{equation}
    
    The variation of the nonminimal action is performed before the vacuum condition is imposed. We will first vary the action with respect to $g_{\mu\nu}$ and $B_\mu$. And then set $X=0$ and solve $B_\mu B^\mu=b^2$ for the radial component of the bumblebee field.
    
    In this paper two cases of potentials will be discussed:  \begin{itemize}
    \item \textbf{Case A}: quadratic potentail with zero cosmological constant
    \begin{equation}
    \label{quadratic potential}
    V=\frac{\lambda X^2}{2},
    \end{equation}
    both $V$ and $V'_X$ vanish at $X=0$. The potential does not appear explicitly in the reduced equations.
    \item \textbf{Case B}: linear potential with none-zero cosmological constant
    \begin{equation}
    \label{linear potential}
    V=\frac{\lambda X}{2},    
    \end{equation}
    in this case the potential vanishes on shell, but $V'_X=\lambda/2$ does not. The potential therefore still contributes to the equations of motion. 
    \end{itemize}
    
    \section{Planar black hole}
    In this section we will discuss method of obtaining exact planar black hole solution in Bumblebee model with none-trivial temporal and radial component of Bumblebee field. 
    We will use the same steps in each case. First, imposed vacuum condition gives the radial component of $B_\mu$. We then combine the bumblebee and gravitational equations. The equations left after integration give the exact form of blackening function and relations between the parameters. Found solution is regular, this solutions being singularity free at the $r=0$. The horizon structure is discussed. 
    
    Let us consider the planar black hole with the metric ansatz
    \begin{equation}
    ds^2=-f(r)\,dt^2+\frac{dr^2}{f(r)}
    +h(r)\left(dx_1^2+dx_2^2\right),
    \label{eq:planar-A-metric}
    \end{equation}
    and the action \eqref{action} with the quadratic potential \eqref{quadratic potential} and zero cosmological constant 
    \begin{equation}
         V(X)=\frac{k}{2}X^2, \qquad \Lambda=0.
    \end{equation}with the temporal and radial components of bumblebee field,
    \begin{equation}
    B_\mu=\left(b_t(r),b_r(r),0,0\right).
    \label{eq:planar-A-bumblebee}
    \end{equation}
    
    Now, we will solve this system by imposing vacuum condition \eqref{vacuumcondition}, $B^\mu B_\mu=b^2$ and get
    \begin{equation}
    -\frac{b_t(r)^2}{f(r)}+f(r)b_r(r)^2=b^2.
    \label{eq:planar-A-vacuum}
    \end{equation}
    Therefore, the radial component of the bumblebee field can be expressed as
    \begin{equation}
    b_r(r)=
    \frac{\sqrt{b_t(r)^2+b^2f(r)}}{f(r)}.
    \label{eq:planar-A-br}
    \end{equation}
    After substituting Eq.~\eqref{eq:planar-A-br} into equation of motion, once can get the system that has three independent gravitational equations
    $\mathcal{E}_{tt}$, $\mathcal{E}_{rr}$, and
    $\mathcal{E}_{xx}$, together with the temporal and radial components of the bumblebee field equation. The simplified system of equation of motion can be written as
    \begin{equation}
    \begin{split}
    \mathcal{E}_{tt}={}&\frac{h(r)b_t'(r)^2}{4b^2\xi f(r)}+\left(\frac{b_t(r)b_t'(r)}{b^2f(r)}+\frac{\left(\xi b_t(r)^2+\left(1+b^2\xi\right)f(r)\right)f'(r)}{2b^2\xi f(r)^2}\right)h'(r)\\
    &-\frac{\left(\xi b_t(r)^2+\left(1+b^2\xi\right)f(r)\right)h'(r)^2}{4b^2\xi f(r)h(r)}+\frac{b_t(r)^2h(r)f''(r)}{2b^2f(r)^2}\\
    &+\frac{\left(\xi b_t(r)^2+\left(1+b^2\xi\right)f(r)\right)h''(r)}{b^2\xi f(r)}=0.
    \end{split}
    \label{eq:planar-A-Ett}
    \end{equation}
    
    \begin{equation}
    \begin{split}
    \mathcal{E}_{rr}={}&\frac{h(r)b_t'(r)^2}{4\xi\left(b_t(r)^2+b^2f(r)\right)}+\left(\frac{b_t(r)b_t'(r)}{b_t(r)^2+b^2f(r)}-\frac{\left(\xi b_t(r)^2-f(r)\right)f'(r)}{2\xi f(r)\left(b_t(r)^2+b^2f(r)\right)}\right)h'(r)\\
    &+\frac{\left(3\xi b_t(r)^2+\left(1+3b^2\xi\right)f(r)\right)h'(r)^2}{4\xi\left(b_t(r)^2+b^2f(r)\right)h(r)}-\frac{h(r)f''(r)}{2f(r)}-h''(r)=0.
    \end{split}
    \label{eq:planar-A-Err}
    \end{equation}
    
    \begin{equation}
    \begin{split}
    \mathcal{E}_{xx}={}&\frac{\left(4\xi-1\right)h(r)b_t'(r)^2}{2b^2\xi f(r)}+\left(\frac{2b_t(r)b_t'(r)}{b^2f(r)}+\frac{\left(1+b^2\xi\right)f'(r)}{b^2\xi f(r)}\right)h'(r)\\
    &-\frac{\left(\xi b_t(r)^2+\left(1+b^2\xi\right)f(r)\right)h'(r)^2}{2b^2\xi f(r)h(r)}+\frac{2b_t(r)h(r)b_t''(r)}{b^2f(r)}\\
    &+\frac{\left(1+b^2\xi\right)h(r)f''(r)}{b^2\xi f(r)}+\frac{\left(\xi b_t(r)^2+\left(1+b^2\xi\right)f(r)\right)h''(r)}{b^2\xi f(r)}=0.
    \end{split}
    \label{eq:planar-A-Exx}
    \end{equation}
    
    The temporal $\mathfrak{B}_{tt}$ and radial $\mathfrak{B}_{rr}$ component of the bumblebee field equations is
    \begin{equation}
    \mathfrak{B}_{tt}=-\frac{b_t'(r)h'(r)}{h(r)}+\frac{\xi b_t(r)f'(r)h'(r)}{f(r)h(r)}+\frac{\xi b_t(r)f''(r)}{f(r)}-b_t''(r)=0.
    \label{eq:planar-A-Bt}
    \end{equation}
    
    \begin{equation}
    \mathfrak{B}_{rr}=-\frac{f'(r)h'(r)}{h(r)}+\frac{f(r)h'(r)^2}{h(r)^2}-\frac{2f(r)h''(r)}{h(r)}-f''(r)=0.
    \label{eq:planar-A-Br}
    \end{equation}
    
    One can solve this system in the following way. From the radial bumblebee equation $\mathfrak{B}_{rr}$ \eqref{eq:planar-A-Br}, one can find
    \begin{equation}
    f''(r)=
    -\frac{f'(r)h'(r)}{h(r)}
    +\frac{f(r)h'(r)^2}{h(r)^2}
    -\frac{2f(r)h''(r)}{h(r)}.
    \label{eq:planar-A-fpp}
    \end{equation}
    
    By substituting \eqref{eq:planar-A-fpp} into the combination
    $\mathcal{E}_{tt}-\mathcal{E}_{rr}$, one obtains
    \begin{equation}
    2\left(1+b^2\xi\right)f(r)
    \left(
    2h(r)h''(r)-h'(r)^2
    \right)=0.
    \label{eq:planar-A-h-equation-full}
    \end{equation}
    For the nondegenerate branch $f(r)\neq0$,
    $1+b^2\xi\neq0$,
    this equation reduces to
    \begin{equation}
    2h(r)h''(r)-h'(r)^2=0.
    \label{eq:planar-A-h-equation}
    \end{equation}
    The solution for the planar deformation function is
    \begin{equation}
    h(r)=\frac{\left(C_1r+2C_2\right)^2}{4C_2}.
    \label{eq:planar-A-h}
    \end{equation}
    
    Substituting Eq.~\eqref{eq:planar-A-h} into
    Eq.~\eqref{eq:planar-A-fpp}, one obtains
    \begin{equation}
    f''(r)
    +\frac{2C_1}{C_1r+2C_2}f'(r)=0.
    \label{eq:planar-A-f-equation}
    \end{equation}
    The solution for the blackening function is
    \begin{equation}
    f(r)=C_4
    -\frac{C_3}{C_1\left(C_1r+2C_2\right)}.
    \label{eq:planar-A-f-general}
    \end{equation}
    
    Using Eqs.~\eqref{eq:planar-A-h} and
    \eqref{eq:planar-A-f-general} in the temporal bumblebee equation
    \eqref{eq:planar-A-Bt}, one obtains
    \begin{equation}
    b_t''(r)
    +\frac{2C_1}{C_1r+2C_2}b_t'(r)=0.
    \label{eq:planar-A-bt-equation}
    \end{equation}
    The solution for the temporal component of the bumblebee field is
    \begin{equation}
    b_t(r)=
    \alpha+\frac{\beta}{C_1r+2C_2}.
    \label{eq:planar-A-bt}
    \end{equation}
    
    Substituting Eqs.~\eqref{eq:planar-A-h},
    \eqref{eq:planar-A-f-general}, and
    \eqref{eq:planar-A-bt} into
    $\mathcal{E}_{x_1x_1}=0$ equation \label{eq:planar-A-Exx}, one obtains
    \begin{equation}
    \left(4\xi-1\right)\beta^2=0.
    \label{eq:planar-A-branch-condition}
    \end{equation}
    For the nonconstant temporal component, $\beta\neq0$, the nonminimal coupling is fixed as
    \begin{equation}
    \xi=\frac14.
    \label{eq:planar-A-xi}
    \end{equation}
    
    Finally, the remaining radial gravitational equation \eqref{eq:planar-A-Err} gives
    \begin{equation}
    C_4=-\frac{\alpha^2}{4+b^2}.
    \label{eq:planar-A-C4}
    \end{equation}
    Now, the radial component of the Bumlebee field will have form 
    \begin{equation}
    b_r(r)=\frac{\sqrt{\left(\alpha+\frac{\beta}{C_1r+2C_2}\right)^2+b^2f(r)}}{f(r)}.
    \label{eq:planar-A-btr}
    \end{equation}
    
    We will demand that $\left(\alpha+\frac{\beta}{C_1r+2C_2}\right)^2+b^2f(r)>0$ for real positive parameters $\alpha$, $\beta$, and $C_3$, so it will give 
    \begin{equation}
        C_2 \geq 0,
        \qquad
        C_1 \geq \frac{b^2C_3}{2\alpha\beta}>0.
    \end{equation}
    The square root function can be rewritten as
    \begin{equation}
    P(r)=\frac{4\alpha^2}{4+b^2}+\frac{2\alpha\beta-\dfrac{b^2C_3}{C_1}}{rC_1+2C_2}+\frac{\beta^2}{\left(rC_1+2C_2\right)^2}.
    \end{equation}
    A particularly simple choice for positively define branch $P(r)$ is
    \begin{equation}
        C_1=\frac{b^2C_3}{2\alpha\beta}, \qquad C_2\geq0.
    \end{equation}
    
    Thus, the exact planar solution takes the form
    \begin{equation}
    f(r)=-\frac{\alpha^{2}}{4+b^{2}}-\frac{\alpha\beta}{b^{2}(2C_{2}+\dfrac{b^{2}rC_{3}}{\alpha\beta})}.
    \label{eq:planar-A-solution}
    \end{equation}
    \begin{equation}
    h(r)=\frac{(2\alpha\beta C_{2}+b^{2}rC_{3})^{2}}{4\alpha^{2}\beta^{2}C_{2}}.
    \end{equation}
    \begin{equation}
    b_{t}(r)=\alpha+\frac{b^{2}C_{3}}{\alpha(\dfrac{2b^{2}C_{2}C_{3}}{\alpha\beta}+\dfrac{b^{4}rC_{3}^{2}}{\alpha^{2}\beta^{2}})}.
    \end{equation}
    Direct substitution of this answer into the three independent gravitational equations and the two bumblebee equations gives zero identically. The vacuum condition \eqref{eq:planar-A-vacuum} is also satisfied identically.
    
    The nonconstant branch assumes $\alpha\neq0, \beta\neq0, C_1\neq0, C_2\neq0.$
    For $\beta=0$, Eq.~\eqref{eq:planar-A-branch-condition} does not fix $\xi$, and the constant-$b_t$ branch must be considered separately.
    
    \subsection{Horizon structure and regularity}
    
    The horizon is determined by the zero of the blackening function $f(r_h)=0$.
    
    Using the solution \eqref{eq:planar-A-solution}, the horizon can be expressed as
    \begin{equation}
    r_h=-\frac{(4+b^{2})\beta^{2}+2b^{2}\alpha\beta C_{2}}{b^{4}C_{3}}.
    \end{equation}
    Moreover, $f'(r_h)=\frac{b^{4}\alpha^{2}\beta^{2}C_{3}}{(-4\beta^{2}-b^{2}\beta^{2})^{2}}$. Which is nonzero for $C_3\neq0$. Therefore, the solution has a regular nonextremal horizon.
    
    We now examine the behavior of the geometry at the origin. For this purpose, we consider the Ricci scalar $R$ and the Kretschmann scalar $K=R_{\mu\nu\rho\sigma}R^{\mu\nu\rho\sigma}$. For out planar black hole it will be 
    \begin{equation}
    R=\frac{2b^{4}\alpha^{2}C_{3}^{2}}{(4+b^{2})(2\alpha\beta C_{2}+b^{2}rC_{3})^{2}},
    \end{equation}
    in the limit $r\to0$ 
    \begin{equation}
    \lim_{r\to 0}R=\frac{b^{4}C_{3}^{2}}{2(4+b^{2})\beta^{2}C_{2}^{2}},
    \end{equation}
    Ricci scalar stays finite, without divergence.
    The limit of Kretschmann scalar is
    \begin{equation}
    \lim_{r\to 0 }K=
    \frac{b^{4}(3(4+b^{2})^{2}\beta^{4}+4b^{2}(4+b^{2})\alpha\beta^{3}C_{2}+4b^{4}\alpha^{2}\beta^{2}C_{2}^{2})C_{3}^{4}}{16(4+b^{2})^{2}\alpha^{2}\beta^{6}C_{2}^{6}}.
    \end{equation}
    Both curvature invariants remain finite for \(C_2\neq0\). Thus, the central curvature singularity is absent. Therefore this solution \eqref{eq:planar-A-solution} describes a regular planar black hole obtained in bumblebee gravity.

    Let us analyze the blackening function, where for the simplicity parameters were chosen as $\alpha=b=C_2=1$, $\beta=C_3=-1$.
    
    \begin{figure}[H]
    \centering
    \includegraphics[width=0.8\textwidth]{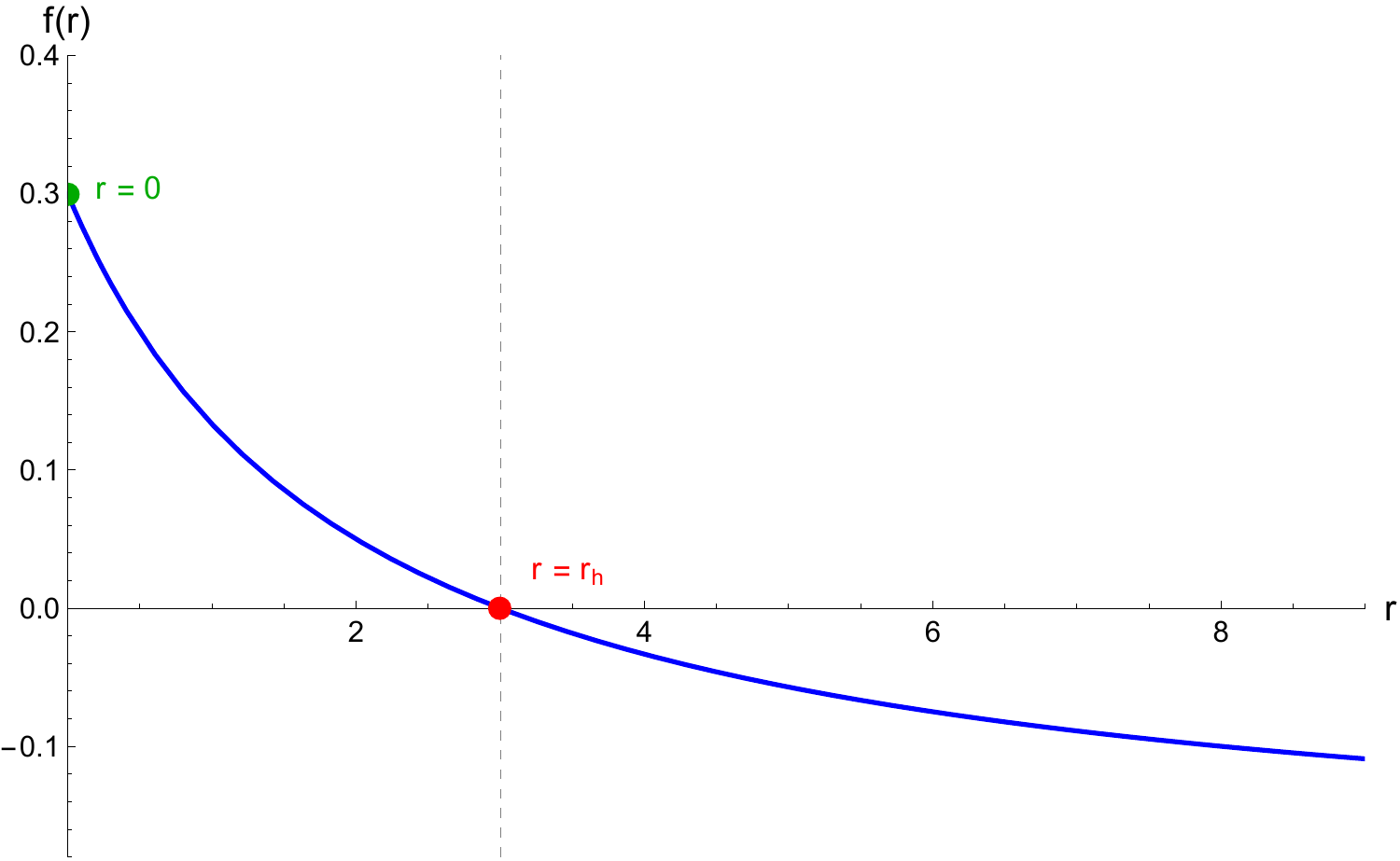}
    \caption{The blackening function for $\alpha=b=C_2=1$ and $\beta=C_3=-1$. The green point denotes $r=0$, while the red point denotes the horizon $r=r_h=3$.}
    \label{fig:planar-A-blackening-regular}
    \end{figure}
    
    The blackening function is finite at $r=0$ and vanishes at $r_h=3$. It is positive for $0\leq r<r_h$ and negative for $r>r_h$. The finite positive value of $f(r)$ at $r=0$ indicates a regular static core. This behavior is similar to that of the de Sitter static patch: the blackening function is positive at $r=0$ and vanishes at a finite cosmological-type horizon.

    \section{Planar black hole with cosmological constant}
    \label{sec:planar-B}
    
    In this section we consider the action \eqref{action} with the linear potential \eqref{linear potential} and nonzero cosmological constant,
    \begin{equation}
    V(X)=\frac{\lambda}{2}X,
    \qquad
    \Lambda\neq0.
    \label{eq:planar-B-potential}
    \end{equation}
    On the vacuum surface $X=0$, the potential vanishes, but its first derivative remains nonzero,
    \begin{equation}
    V(X)=0,
    \qquad
    V'_X=\frac{\lambda}{2}.
    \end{equation}
    Therefore, the potential contributes explicitly to the equations of motion.
    
    We consider the planar metric in the form \eqref{eq:planar-A-metric},
    with bumblebee field ansatz \eqref{eq:planar-A-bumblebee} for temporal and radial components of the  as before.
    
    From the vacuum condition \eqref{vacuumcondition}, one obtains
    \begin{equation}
    -\frac{b_t(r)^2}{f(r)}+f(r)b_r(r)^2=b^2.
    \label{eq:planar-B-vacuum}
    \end{equation}
    The radial component of the bumblebee field can therefore be expressed as
    \begin{equation}
    b_r(r)=
    \frac{\sqrt{b_t(r)^2+b^2f(r)}}{f(r)}.
    \label{eq:planar-B-br}
    \end{equation}
    
    After substituting Eq.~\eqref{eq:planar-B-br} into the equations of motion, one obtains three independent gravitational equations
    $\mathcal{E}_{tt}$, $\mathcal{E}_{rr}$, and $\mathcal{E}_{xx}$, together with the temporal and radial components of the bumblebee field equation. The simplified system can be written as
    \begin{equation}
    \begin{split}
    \mathcal{E}_{tt}={}&
    \frac{\left(\lambda b_t(r)^2+2\Lambda f(r)\right)h(r)}
    {2b^2\xi f(r)^2}
    +\frac{h(r)b_t'(r)^2}{4b^2\xi f(r)}
    \\
    &+\left(
    \frac{b_t(r)b_t'(r)}{b^2f(r)}
    +\frac{\left(\xi b_t(r)^2+\left(1+b^2\xi\right)f(r)\right)f'(r)}
    {2b^2\xi f(r)^2}
    \right)h'(r)
    \\
    &-\frac{\left(\xi b_t(r)^2+\left(1+b^2\xi\right)f(r)\right)h'(r)^2}
    {4b^2\xi f(r)h(r)}
    +\frac{b_t(r)^2h(r)f''(r)}{2b^2f(r)^2}
    \\
    &+\frac{\left(\xi b_t(r)^2+\left(1+b^2\xi\right)f(r)\right)h''(r)}
    {b^2\xi f(r)}=0.
    \end{split}
    \label{eq:planar-B-Ett}
    \end{equation}
    
    \begin{equation}
    \begin{split}
    \mathcal{E}_{rr}={}&
    -\frac{\left(\lambda b_t(r)^2-2\Lambda f(r)+b^2\lambda f(r)\right)h(r)}
    {2\xi f(r)\left(b_t(r)^2+b^2f(r)\right)}
    +\frac{h(r)b_t'(r)^2}
    {4\xi\left(b_t(r)^2+b^2f(r)\right)}
    \\
    &+\left(
    \frac{b_t(r)b_t'(r)}{b_t(r)^2+b^2f(r)}
    -\frac{\left(\xi b_t(r)^2-f(r)\right)f'(r)}
    {2\xi f(r)\left(b_t(r)^2+b^2f(r)\right)}
    \right)h'(r)
    \\
    &+\frac{\left(3\xi b_t(r)^2+\left(1+3b^2\xi\right)f(r)\right)h'(r)^2}
    {4\xi\left(b_t(r)^2+b^2f(r)\right)h(r)}
    -\frac{h(r)f''(r)}{2f(r)}
    -h''(r)=0.
    \end{split}
    \label{eq:planar-B-Err}
    \end{equation}
    
    \begin{equation}
    \begin{split}
    \mathcal{E}_{xx}={}&
    \frac{2\Lambda h(r)}{b^2\xi f(r)}
    +\frac{\left(4\xi-1\right)h(r)b_t'(r)^2}{2b^2\xi f(r)}
    \\
    &+\left(
    \frac{2b_t(r)b_t'(r)}{b^2f(r)}
    +\frac{\left(1+b^2\xi\right)f'(r)}{b^2\xi f(r)}
    \right)h'(r)
    \\
    &-\frac{\left(\xi b_t(r)^2+\left(1+b^2\xi\right)f(r)\right)h'(r)^2}
    {2b^2\xi f(r)h(r)}
    +\frac{2b_t(r)h(r)b_t''(r)}{b^2f(r)}
    \\
    &+\frac{\left(1+b^2\xi\right)h(r)f''(r)}{b^2\xi f(r)}
    +\frac{\left(\xi b_t(r)^2+\left(1+b^2\xi\right)f(r)\right)h''(r)}
    {b^2\xi f(r)}=0.
    \end{split}
    \label{eq:planar-B-Exx}
    \end{equation}
    
    The temporal $\mathfrak{B}_{tt}$ and radial $\mathfrak{B}_{rr}$ components of the bumblebee field equations are
    \begin{equation}
    \begin{split}
    \mathfrak{B}_{tt}={}&
    \frac{\lambda b_t(r)}{f(r)}
    -\frac{b_t'(r)h'(r)}{h(r)}
    +\frac{\xi b_t(r)f'(r)h'(r)}{f(r)h(r)}
    \\
    &+\frac{\xi b_t(r)f''(r)}{f(r)}
    -b_t''(r)=0.
    \end{split}
    \label{eq:planar-B-Bt}
    \end{equation}
    
    \begin{equation}
    \mathfrak{B}_{rr}=
    -\frac{\lambda}{\xi}
    -\frac{f'(r)h'(r)}{h(r)}
    +\frac{f(r)h'(r)^2}{h(r)^2}
    -\frac{2f(r)h''(r)}{h(r)}
    -f''(r)=0.
    \label{eq:planar-B-Br}
    \end{equation}
    
    One can solve this system in the following way. From the radial bumblebee equation $\mathfrak{B}_{rr}$ \eqref{eq:planar-B-Br}, one can find
    \begin{equation}
    f''(r)=
    -\frac{\lambda}{\xi}
    -\frac{f'(r)h'(r)}{h(r)}
    +\frac{f(r)h'(r)^2}{h(r)^2}
    -\frac{2f(r)h''(r)}{h(r)}.
    \label{eq:planar-B-fpp}
    \end{equation}
    
    By substituting Eq.~\eqref{eq:planar-B-fpp} into the combination of the temporal and radial gravitational equations $\mathcal{E}_{tt}
    -\mathcal{E}_{rr}$, one obtains
    \begin{equation}
    -\frac{\left(1+b^2\xi\right)f(r)
    \left(h'(r)^2-2h(r)h''(r)\right)}
    {2\xi\left(b_t(r)^2+b^2f(r)\right)h(r)}
    =0.
    \label{eq:planar-B-h-equation-full}
    \end{equation}
    For the nondegenerate branch $f(r)\neq0$ and $1+b^2\xi\neq0$, this equation reduces to
    \begin{equation}
    2h(r)h''(r)-h'(r)^2=0.
    \label{eq:planar-B-h-equation}
    \end{equation}
    The solution for the planar deformation function is
    \begin{equation}
    h(r)=
    \frac{\left(C_1r+2C_2\right)^2}{4C_2}.
    \label{eq:planar-B-h}
    \end{equation}
    
    Substituting Eq.~\eqref{eq:planar-B-h} into Eq.~\eqref{eq:planar-B-fpp}, one obtains
    \begin{equation}
    f''(r)
    +\frac{2C_1}{C_1r+2C_2}f'(r)
    +\frac{\lambda}{\xi}=0.
    \label{eq:planar-B-f-equation}
    \end{equation}
    The solution for the blackening function is
    \begin{equation}
    f(r)=
    C_4
    -\frac{\lambda r^2}{6\xi}
    -\frac{2\lambda rC_2}{3\xi C_1}
    -\frac{C_3}{C_1\left(C_1r+2C_2\right)}.
    \label{eq:planar-B-f-general}
    \end{equation}
    
    Using Eqs.~\eqref{eq:planar-B-h} and \eqref{eq:planar-B-f-general} in the temporal bumblebee equation \eqref{eq:planar-B-Bt}, one obtains
    \begin{equation}
    b_t''(r)
    +\frac{2C_1}{C_1r+2C_2}b_t'(r)=0.
    \label{eq:planar-B-bt-equation}
    \end{equation}
    The solution for the temporal component of the bumblebee field is
    \begin{equation}
    b_t(r)=
    \alpha+\frac{\beta}{C_1r+2C_2}.
    \label{eq:planar-B-bt}
    \end{equation}
    
    Substituting Eqs.~\eqref{eq:planar-B-h}, \eqref{eq:planar-B-f-general}, and \eqref{eq:planar-B-bt} into the gravitational equation $\mathcal{E}_{xx}=0$ \eqref{eq:planar-B-Exx}, one obtains the relation
    \begin{equation}
    -2\xi\Lambda+\lambda+b^2\xi\lambda=0.
    \label{eq:planar-B-Lambda-condition}
    \end{equation}
    Therefore, the cosmological constant is related to the potential parameter as
    \begin{equation}
    \Lambda=
    \frac{\lambda\left(1+b^2\xi\right)}{2\xi}.
    \label{eq:planar-B-Lambda-general}
    \end{equation}
    
    Using Eq.~\eqref{eq:planar-B-Lambda-general} in the same gravitational equation, the remaining condition is
    \begin{equation}
    \left(4\xi-1\right)\beta^2=0.
    \label{eq:planar-B-branch-condition}
    \end{equation}
    For the nonconstant temporal component, $\beta\neq0$, the nonminimal coupling is fixed as
    \begin{equation}
    \xi=\frac14.
    \label{eq:planar-B-xi}
    \end{equation}
    Consequently, Eq.~\eqref{eq:planar-B-Lambda-general} gives
    \begin{equation}
    \Lambda=\frac{4+b^2}{2}\lambda.
    \label{eq:planar-B-Lambda}
    \end{equation}
    
    Finally, the remaining radial gravitational equation \eqref{eq:planar-B-Err} gives
    \begin{equation}
    C_4=
    -\frac{\alpha^2}{4+b^2}
    -\frac{8\lambda C_2^2}{3C_1^2}.
    \label{eq:planar-B-C4}
    \end{equation}
    
    Thus, the radial component is 
    \begin{equation}
    b_r(r)=
    \frac{\sqrt{
    \left(\alpha+\frac{\beta}{C_1r+2C_2}\right)^2+b^2f(r)
    }}{f(r)}.
    \label{eq:planar-B-br-final}
    \end{equation}
    
    and blackening function 
    \begin{equation}
    f(r)=
    -\frac{\alpha^2}{4+b^2}
    -\frac{2\lambda\left(C_1r+2C_2\right)^2}{3C_1^2}
    -\frac{C_3}{C_1\left(C_1r+2C_2\right)}.
    \label{eq:planar-B-solution}
    \end{equation}.
    
    The nonconstant branch assumes $\alpha\neq0$, $\beta\neq0$, $C_1\neq0$, $C_2\neq0$, and $\lambda\neq0$. The bumblebee field is real in the region where
    \begin{equation}
    \left(\alpha+\frac{\beta}{C_1r+2C_2}\right)^2+b^2f(r)\geq0.
    \label{eq:planar-B-reality}
    \end{equation}
    
    We now discuss the condition ensuring that the radial component of the bumblebee field \eqref{eq:planar-B-br-final} is real. The expression under the square root can be written as
    \begin{equation}
    \begin{split}
    b_t(r)^2+b^2f(r)={}&
    \frac{4\alpha^2}{4+b^2}
    +\frac{2\alpha\beta-\dfrac{b^2C_3}{C_1}}{C_1r+2C_2}
    +\frac{\beta^2}{\left(C_1r+2C_2\right)^2}
    \\
    &-\frac{2b^2\lambda}{3C_1^2}
    \left(C_1r+2C_2\right)^2.
    \end{split}
    \label{eq:planar-B-square-root}
    \end{equation}
    
    For fixed real parameters $\alpha>0$, $\beta>0$, $C_3>0$, and $b\neq0$, a sufficient condition ensuring positivity for every $r>0$ is
    \begin{equation}
    \lambda\leq0,
    \qquad
    C_2\geq0,
    \qquad
    C_1\geq\frac{b^2C_3}{2\alpha\beta}>0.
    \label{eq:planar-B-positivity}
    \end{equation}
    Indeed, these conditions give $C_1r+2C_2>0$ for every $r>0$, and all terms on the right-hand side of Eq.~\eqref{eq:planar-B-square-root} are nonnegative. Since the first term is strictly positive, one obtains
    \begin{equation}
    b_t(r)^2+b^2f(r)>0.
    \end{equation}
    
     Moreover, the present case assumes $\Lambda\neq0$, which requires $\lambda\neq0$ by Eq.~\eqref{eq:planar-B-Lambda}. Therefore, the sufficient physical branch is
    \begin{equation}
    \lambda<0,
    \qquad
    C_2>0,
    \qquad
    C_1\geq\frac{b^2C_3}{2\alpha\beta}>0.
    \label{eq:planar-B-physical-branch}
    \end{equation}
    For this branch, the cosmological constant is negative,
    \begin{equation}
    \Lambda=\frac{4+b^2}{2}\lambda<0.
    \end{equation}
    
    As in the case without cosmological constant, a particularly simple choice is
    \begin{equation}
    C_1=\frac{b^2C_3}{2\alpha\beta},
    \qquad
    C_2>0,
    \qquad
    \lambda<0.
    \label{eq:planar-B-simple-choice}
    \end{equation}
    For this choice, the term proportional to $\left(C_1r+2C_2\right)^{-1}$ in Eq.~\eqref{eq:planar-B-square-root} vanishes, and the expression under the square root becomes
    \begin{equation}
    b_t(r)^2+b^2f(r)=
    \frac{4\alpha^2}{4+b^2}
    +\frac{\beta^2}{\left(C_1r+2C_2\right)^2}
    -\frac{2b^2\lambda}{3C_1^2}
    \left(C_1r+2C_2\right)^2>0.
    \label{eq:planar-B-square-root-positive}
    \end{equation}
    Thus, the square root remains real and strictly positive for every $r>0$, and its value at $r=0$ is also finite and positive.
    
    The blackening and planar function in this case will be 
    
    \begin{equation}
    f(r)=-\frac{\alpha^{2}}{4+b^{2}}-\frac{2r^{2}\lambda}{3}-\frac{8\alpha^{2}\beta^{2}\lambda C_{2}^{2}}{3b^{4}C_{3}^{2}}-\frac{8r\alpha\beta\lambda C_{2}}{3b^{2}C_{3}}-\frac{\alpha^{2}\beta^{2}}{2b^{2}\alpha\beta C_{2}+b^{4}rC_{3}}.
    \label{eq:planar-B-solution}
    \end{equation}
    \begin{equation}
    h(r)=\frac{(2C_{2}+\dfrac{b^{2}rC_{3}}{\alpha\beta})^{2}}{4C_{2}}.
    \end{equation}
    The parameters satisfy
    \begin{equation}
    \xi=\frac14,
    \qquad
    \Lambda=\frac{4+b^2}{2}\lambda.
    \label{eq:planar-B-parameters}
    \end{equation}
    
    For $\beta=0$, Eq.~\eqref{eq:planar-B-branch-condition} does not fix $\xi$, and the constant-$b_t$ branch must be considered separately.
    
    \subsection{Regularity and horizon structure}
    
    We first examine the behavior of the geometry at the origin. For this purpose, we consider the Ricci scalar $R$ and the Kretschmann scalar $K=R_{\mu\nu\rho\sigma}R^{\mu\nu\rho\sigma}$.
    
    For the choice $C_1=\frac{b^2C_3}{\alpha\beta}$ used in the following expressions, the limit of Ricci scalar $r\to0$ is 
    \begin{equation}
    \lim_{r\to0}R=
    \frac{
    32(4+b^2)\alpha^2\beta^2\lambda C_2^2
    +2b^4\alpha^2C_3^2
    }{
    4(4+b^2)\alpha^2\beta^2C_2^2
    }.
    \end{equation}
    The Ricci scalar stays finite, without divergence.
    
    The limit of the Kretschmann scalar is
    \begin{equation}
    \begin{split}
    \lim_{r\to0}K={}&
    \frac{1}{48(4+b^2)^2\alpha^6\beta^6C_2^6}
    \Bigl(
    9b^4(4+b^2)^2\alpha^4\beta^4C_3^4
    \\
    &\qquad
    +12b^6(4+b^2)\alpha^5\beta^3C_2C_3^4
    \\
    &\qquad
    +4\alpha^6\beta^2C_2^2
    \Bigl(
    128(4+b^2)^2\beta^4\lambda^2C_2^4
    \\
    &\qquad\qquad
    +16b^4(4+b^2)\beta^2\lambda C_2^2C_3^2
    +3b^8C_3^4
    \Bigr)
    \Bigr).
    \end{split}
    \end{equation}
    
    Both curvature invariants remain finite for nonzero $\alpha$, $\beta$, and $C_2$. Thus, there is no curvature singularity at $r=0$. Moreover, for positive parameters $\alpha$, $\beta$, $C_2$, and $C_3$, with $b\neq0$, the denominator $2\alpha\beta C_2+b^2rC_3$ does not vanish for $r\geq0$. Therefore, the curvature invariants remain finite throughout this radial domain.
    
    We now discuss the horizon structure. The horizon is determined by the zero of the blackening function $f(r_h)=0$.
    
    Using the solution \eqref{eq:planar-B-solution} with the same choice of $C_1$, the horizon equation takes the form
    \begin{equation}
    \begin{split}
    &2\lambda(4+b^2)(2\alpha\beta C_2+b^2r_hC_3)^3
    \\
    &+3b^4\alpha^2C_3^2(2\alpha\beta C_2+b^2r_hC_3)
    +3b^2(4+b^2)\alpha^2\beta^2C_3^2=0.
    \end{split}
    \end{equation}
    The horizon $r_h$ can be determined as the positive real root of this cubic equation.
    
    For the chosen positive parameters and $\lambda<0$, the blackening function is strictly increasing and tends to $+\infty$ as $r\to\infty$. Therefore, a unique positive horizon exists when
    \begin{equation}
    f(0)=
    -\frac{\alpha^2}{4+b^2}
    -\frac{\alpha\beta}{2b^2C_2}
    -\frac{8\alpha^2\beta^2\lambda C_2^2}{3b^4C_3^2}<0.
    \end{equation}
    Moreover,
    \begin{equation}
    f'(r_h)=
    -\frac{4\lambda(2\alpha\beta C_2+b^2r_hC_3)}{3b^2C_3}
    +\frac{\alpha^2\beta^2C_3}{(2\alpha\beta C_2+b^2r_hC_3)^2}>0.
    \end{equation}
    Thus, the horizon is nonextremal. Both curvature invariants remain finite at $r=r_h$, and the zero of the blackening function corresponds to a regular metric horizon.
    
    The blackening function is negative for $0\leq r<r_h$ and positive for $r>r_h$. 
    
    \section{Conclusions}
    \label{sec:conclusions}
In this paper, we have presented two families of exact planar black hole solutions in bumblebee gravity with nontrivial temporal and radial components of the bumblebee field. We considered a quadratic potential without a cosmological constant and a linear potential with a nonzero cosmological constant.

We obtained conditions ensuring a real bumblebee field and analyzed the regularity and horizon structure of the solutions. The curvature invariants remain finite both at the origin and at the horizon for the parameter domains considered. The solution without a cosmological constant exhibits a cosmological-type horizon.

These solutions show that nonminimally coupled Lorentz-violating vector field or pure Bumblebee fields can supports planar geometries without a central scalar-curvature singularity. The thermodynamics and stability of these solutions remain subjects for further study.

    \section*{Acknowledgement}
    This paper is dedicated to the loving memory of Oleg Stepanenko (10.08.1947-09.11.2013).
    
    We would like to thank F. Karakus, I. Moiseenkov, M. Mullahasanoglu, A. Polevov, P. Slepov, A. Teretenkov, I. Volovich, and J. Wu for useful discussions. I would also like to express my gratitude to E. Menshenina and acknowledge her earlier contribution to this work.
    
    The work of D. Stepanenko was performed at the Steklov International Mathematical Center and supported by the Ministry of Science and Higher Education of the Russian Federation (agreement no. 075-15-2025-303).
    And was supported by the Theoretical Physics and Mathematics Advancement Foundation ``BASIS'' (no. 24-1-1-82-1).

    \end{document}